\documentclass[aps,prb,groupedaddress,twocolumn,amsmath,amssymb]{revtex4}
    \usepackage{dcolumn}
    \usepackage{bm}
    \usepackage{graphicx}
    \usepackage{dcolumn}
\newcommand\beq{\begin{equation}}
\newcommand\eeq{\end{equation}}
\newcommand\beqa{\begin{eqnarray}}
\newcommand\eeqa{\end{eqnarray}}
\newcommand{\nn}{\nonumber\\}

\newcommand{\mo}{\mu_1}
\newcommand{\mt}{\mu_2}
\newcommand{\mth}{\mu_3}
\newcommand{\mn}{\mu_n}
\newcommand{\pure}{{\text{p}}}

\begin{document}



\title{Contact values of the
particle-particle and  wall-particle correlation functions in a
hard-sphere polydisperse fluid}


\author{Andr\'es Santos}
\email{andres@unex.es}
\homepage{http://www.unex.es/eweb/fisteor/andres/}
\author{Santos B. Yuste}
\email{santos@unex.es}
\homepage{http://www.unex.es/eweb/fisteor/santos/}
\affiliation{Departamento de F\'{\i}sica, Universidad de
Extremadura, Badajoz, E-06071, Spain}
\author{Mariano L\'{o}pez de Haro}
\email{malopez@servidor.unam.mx}
\homepage{http://miquiztli.cie.unam.mx/xml/tc/ft/mlh/}
\affiliation{Centro de Investigaci\'on en Energ\'{\i}a, UNAM,
Temixco, Morelos 62580, M{e}xico}

\date{\today}

\begin{abstract}
The contact values $g(\sigma,\sigma')$ of the radial distribution
functions of a  fluid of {(additive)} hard spheres with a given size
distribution $f(\sigma)$ are considered. A ``universality''
assumption is introduced, according to which, at a given packing
fraction $\eta$, $g(\sigma,\sigma')=G(z(\sigma,\sigma'))$, where $G$
is a common function independent of the number of components (either
finite or infinite) and $z(\sigma,\sigma')=[2 \sigma
\sigma'/(\sigma+\sigma')]\mt/\mth$ is a dimensionless parameter,
$\mn$ being the $n$-th moment of the diameter distribution. A cubic
form proposal for the $z$-dependence of $G$ is made and known exact
consistency conditions for the point particle and equal size limits,
as well as between two different routes to compute the pressure of
the system in the presence of a hard wall, are used to express
$G(z)$ in terms of the radial distribution at contact of the
one-component system. For  polydisperse systems we compare the
contact values of the wall-particle correlation function  and the
compressibility factor with those obtained from recent Monte Carlo
simulations.
\end{abstract}

\date{\today}


\maketitle

\section{Introduction}
\label{sec1}

The prominence of hard-core fluids in liquid state theory as
prototype systems for theoretical understanding and stepping stone
for the study of more realistic fluids can hardly be overemphasized.
It is well known that the form of the pressure equation of these
systems acquires a particularly simple representation in terms of
the contact values of the radial distribution functions (rdf).
Therefore, since for hard-core fluids the internal energy reduces to
that of ideal gases, the knowledge of such contact values would be
enough to obtain their equation of state (EOS) and all their
thermodynamic properties.\cite{S05} Unfortunately up to the present
day, and except for the case of hard rods, no exact expressions
either for the contact values of the rdf or for the EOS are
available and so, in order to make progress, research has relied on
various approximate (mostly empirical or semiempirical) approaches
or on the results of simulations. The situation is rather more
complicated for mixtures than for single component fluids and hence
it is not surprising that studies for the former are fewer. From the
analytical point of view, perhaps the most important result is the
exact solution of the Percus--Yevick (PY) equation of additive
hard-sphere mixtures carried out by Lebowitz,\cite{L64} that
includes explicit expressions for the contact values of the rdf. In
turn, such expressions served as the basis for the derivation of the
widely used and rather accurate
Boubl\'{\i}k--Mansoori--Carnahan--Starling--Leland (BMCSL)
EOS\cite{B70,MCSL71} {for} hard-sphere mixtures. In fact,
Boubl\'{\i}k\cite{B70} {(and, independently, Grundke and
Henderson\cite{GH72} and Lee and Levesque\cite{LL73}) proposed an
interpolation between the PY contact values and the ones of the
Scaled Particle Theory (SPT).\cite{LHP65,R88} We will refer to this
interpolation as the Boubl\'{\i}k--Grundke--Henderson--Lee--Levesque
(BGHLL)} approximation for the contact values, refinements of which
have been subsequently introduced, among others, by Henderson
\textit{et al.},\cite{Henderson} Matyushov and Ladanyi,\cite{ML97}
and Barrio and Solana\cite{BS00} to eliminate some drawbacks of the
BMCSL EOS in the so-called colloidal limit of {binary} hard-sphere
mixtures.

It is interesting to point out that in the case of multicomponent
mixtures of hard spheres, the contact values which follow from the
solution of the PY equation, \cite{L64} those of the SPT
approximation,\cite{LHP65,R88} and those of the BGHLL
interpolation\cite{B70,GH72,LL73} present a kind of {``universal''
behavior in the following sense. Once} the packing fraction is
fixed, the expressions for the contact values of the rdf for all
pairs of like and unlike species depend on the diameters of both
species and on the size {distribution} \textit{only} through a
single dimensionless parameter, irrespective of the number of
components in the mixture. In previous work\cite{SYH99,Contact} we
have introduced approximate expressions for the contact values of
the rdf valid for mixtures with an arbitrary number of components
and in arbitrary dimensionality, that require as input the EOS of
the one-component fluid. Apart from satisfying known consistency
conditions,  they are sufficiently general and flexible to
accommodate any given EOS for the single fluid and also share the
universal behavior alluded to above. In the latter
paper,\cite{Contact} two functional forms (a quadratic one and a
rational one) were examined. We found that the best global agreement
with the available simulation results for binary and ternary
mixtures was provided by the quadratic function, which has a
structure similar to the SPT and BGHLL prescriptions, except perhaps
for very disparate mixtures, where the rational approximation seemed
to be preferable.

The universality feature present in the above proposals, which
applies to mixtures with an arbitrary number of components $N$ {and
an arbitrary size distribution}, permits in principle to consider
different situations. For instance, one could study the structural
properties of an $M$-component mixture in the presence of a hard
wall by considering a mixture with $N=M+1$ components and taking the
limit in which the diameter of one of the species goes to infinity.
Also, one could take the limit $N\to \infty$ corresponding to a
polydisperse system of hard spheres in which rather than a discrete
set of values for the diameters, one has a continuous distribution.
Interest in studying the thermodynamic and structural properties of
these polydisperse systems dates back to the late 1970s and the
1980s\cite{others1} and  has recently been revived.\cite{others2} Of
particular concern to us here is a recent paper by Buzzacchi
\textit{et al.}\cite{Ignacio} in which they study the structural
properties of polydisperse hard spheres in the presence of a hard
wall. It seems natural to compare their results with the ones
derived from other
theories,\cite{L64,B70,MCSL71,LHP65,R88,GH72,LL73} in particular
from our approach.\cite{SYH99,Contact} However, as shown below,
except for the SPT contact values (which are known to be generally
less accurate than other proposals), the rest of the approximations
(including our {previous proposals\cite{SYH99,Contact}}) lead to an
inconsistency between two different ways of computing the pressure
in the polydisperse fluid.

The major aim of this paper is to provide yet another (more general)
approximation for the contact values of the rdf that preserves the
property of universality present in the PY, SPT, and BGHLL, as well
as in our {previous approximations}, but avoids the problem just
stated. It is with this new approximation that we will compare the
results of Ref.\ \onlinecite{Ignacio}.

The paper is organized as follows. In Sec.\ \ref{sec2} we derive the
new proposal for the contact values of the rdf using the known
consistency conditions and two different routes to compute the
compressibility factor of the polydisperse hard-sphere system.
Section \ref{sec3} deals with the comparison between our contact
values, the ensuing compressibility factors, and the results of
Buzzacchi \textit{et al.}\cite{Ignacio} and other theories. We close
the paper in Sec.\ \ref{sec4} with further discussion and some
concluding remarks.

\section{Contact values of the radial distribution functions}
\label{sec2}
\subsection{Our proposal}
Let us consider a polydisperse hard-sphere mixture with a given size
distribution $f(\sigma)$ (either {continuous or discrete, the latter
being of the form $f(\sigma)=\sum_i x_i \delta(\sigma_i-\sigma)$})
at a given packing fraction $\eta=\frac{\pi}{6}\rho \mth$, where
$\rho$ is the (total) number density and
\beq
\mn\equiv\langle \sigma^n\rangle=\int_0^\infty d\sigma\, \sigma^n
f(\sigma)
\label{1}
\eeq
denotes the $n$-th moment of the size distribution.

We will use the notation $g(\sigma,\sigma')$ for the contact value
of the pair correlation function of particles of diameters $\sigma$
and $\sigma'$. This function enters into the virial expression of
the EOS as\cite{Lado}
\beqa
Z\equiv\frac{p}{\rho k_B T}&=&1+4\frac{\eta}{\mth}\int_0^\infty
d\sigma\int_0^\infty d\sigma'\, f(\sigma)f(\sigma')\nn
&&\times
\left(\frac{\sigma+\sigma'}{2}\right)^3g(\sigma,\sigma')\nn
&=&1+\frac{\eta}{2
\mth}\left\langle\left(\sigma+\sigma'\right)^3g(\sigma,\sigma')\right\rangle,
\label{2}
\eeqa
where $p$ is the pressure and $T$ is the absolute
temperature.

 A hard wall can
be seen as a sphere of infinite diameter. As a consequence, the
contact value of the correlation function $g_w(\sigma)$ of a sphere
of diameter $\sigma$ with the wall is obtained from
$g(\sigma,\sigma')$ as
\beq
g_w(\sigma)=\lim_{\sigma'\to\infty}g(\sigma,\sigma').
\label{3}
\eeq
An alternative route to the EOS is then provided by using the sum
rule connecting the pressure and the above contact
values,\cite{Evans} namely
\beq
Z_w=\int_0^\infty d\sigma\,f(\sigma)g_w(\sigma)=\langle
g_w(\sigma)\rangle.
\label{4}
\eeq
The subscript $w$ in $Z_w$ has been used to
emphasize that Eq.\ (\ref{4}) represents a route alternative to the
virial one, Eq.\ (\ref{2}), to get the EOS of the hard-sphere
polydisperse fluid. Of course, $Z=Z_w$ in an exact description, but
$Z$ and $Z_w$ may differ when dealing with approximate expressions
for $g(\sigma,\sigma')$ and the associated $g_w(\sigma)$.

Now we consider a class of approximations of the type\cite{Contact}
\beq
g(\sigma,\sigma')=G(z(\sigma,\sigma')),
\label{n1}
\eeq
where
\beq
z(\sigma,\sigma')\equiv
\frac{2\sigma\sigma'}{\sigma+\sigma'}\frac{\mt}{\mth}
\label{7}
\eeq
is a dimensionless parameter depending on the diameters $\sigma$ and
$\sigma'$, as well as on the second and third moments. According to
Eq.\ (\ref{n1}), at a given packing fraction $\eta$ all the
dependence of $g(\sigma,\sigma')$ on $\sigma$, $\sigma'$, and the
details of the size distribution $f(\sigma)$ occurs through the
{single} parameter $z(\sigma,\sigma')$. Once one accepts the
{``univerality''} ansatz (\ref{n1}), it remains to propose an
explicit form for the function $G(z)$. To that end, some consistency
conditions might be useful. First, in the one-component limit,
{\textit{i.e.}, $f(\sigma)=\delta(\sigma-\sigma_0)$}, one has $z=1$,
so that\cite{SYH99,Contact}
\beq
G(z=1)=g_{\text{p}},
\label{n2}
\eeq
where $g_{\text{p}}$ is the contact value of the radial distribution
function of the one-component fluid at the same packing fraction
$\eta$ as the packing fraction of the mixture. Next, the case of a
mixture in which one of the species is made of point particles,
\textit{i.e.}, $\sigma \to 0$, leads to\cite{SYH99,Contact}
\beq
G(z=0)=\frac{1}{1-\eta}\equiv G_0.
\label{6}
\eeq
Conditions (\ref{n2}) and (\ref{6}) are the basic ones. A more
stringent condition is the self-consistency between the routes
(\ref{2}) and (\ref{4}) {for any distribution $f(\sigma)$}. To
proceed further, let us express $G(z)$ as a series in powers of $z$:
\beq
G(z)=G_0+\sum_{n=1}^\infty G_n z^n,
\label{n4}
\eeq
where it has been assumed that $z=0$ is a regular point. Condition
(\ref{6}) is already built in. {In agreement with the universality
assumption (\ref{n1}), the coefficients $G_n$ are independent of the
size distribution, being functions of the packing fraction $\eta$
only}. After simple algebra, the compressibility factor obtained by
inserting the ansatz (\ref{n1}){, along with Eq.\ (\ref{n4}),} into
Eq.\ (\ref{2}) reads
\beqa
Z&=&G_0+3\eta\frac{\mo\mt}{\mth}G_0+\eta\sum_{n=1}^\infty
2^{n-1}G_n\frac{\mt^n}{\mth^{n+1}}\nn &&\times
\langle\sigma^n{\sigma'}^n\left(\sigma+\sigma'\right)^{3-n}\rangle.
\label{n5}
\eeqa

At this point, we impose the condition that this compressibility
factor  depends functionally on the size distribution $f(\sigma)$
only through a finite number of moments. This implies that the
series in Eq.\ (\ref{n5}) must be truncated after $n=3$. Therefore,
we restrict ourselves to the class of  approximations
\beq
G(z)=G_0+G_1 z+G_2 z^2+G_3 z^3.
\label{5}
\eeq
Using the
approximation (\ref{5}) in Eq.\ (\ref{n5}) we get
\beq
Z=G_0+\eta\left[\frac{\mo \mt}{\mth}
\left(3G_0+2G_1\right)+2\frac{\mt^3}{\mth^2}\left(G_1+2G_2+2G_3\right)\right].
\label{11}
\eeq
{Note that the dependence of $Z$ on $f(\sigma)$ through $\mo$,
$\mt$, and $\mth$ is explicit. It only remains to determine the
$\eta$-dependence of $G_1$, $G_2$, and $G_3$.}

Now we turn to the alternative route to derive the compressibility
factor using Eq.\ (\ref{4}). {}From Eqs.\ (\ref{3}), (\ref{n1}), and
(\ref{5}) one obtains the approximation
\beq
g_w(\sigma)=G_0+G_1 z_w(\sigma)+G_2 z_w^2(\sigma)+G_3 z_w^3(\sigma),
\label{12} \eeq where \beq
z_w(\sigma)=\lim_{\sigma'\to\infty}z(\sigma,\sigma')=
{2\sigma}\frac{\mt}{\mth}.
\label{13}
\eeq
Thus the EOS (\ref{4})
then becomes
\beq Z_w=G_0+2\frac{\mo \mt}{\mth}G_1
+4\frac{\mt^3}{\mth^2}\left(G_2+2G_3\right).
\label{14}
\eeq
{Again, the dependence of $Z_w$ on the distribution moments is
explicit. In fact, both $Z$ and $Z_w$ are linear in the combinations
$\mo\mt/\mth$ and $\mt^3/\mth^2$.} The difference between Eqs.\
(\ref{11}) and (\ref{14}) is given by
\beqa
Z-Z_w&=&\frac{\mo \mt}{\mth}\left[3\eta G_0-2(1-\eta)G_1\right]
+2\frac{\mt^3}{\mth^2}\left[\eta G_1\right.\nn
&&\left.-2(1-\eta)G_2-2(2-\eta)G_3\right].
\label{15}
\eeqa

If we want to have $Z=Z_w$ for any dispersity, the coefficients of
${\mo \mt}/{\mth}$ and of ${\mt^3}/{\mth^2}$ in Eq.\ (\ref{15}) must
vanish simultaneously. This gives
\beq
G_1=\frac{3 \eta}{2
\left(1-\eta\right)^2},
\label{n7a}
\eeq
and
\beq G_2=\frac{3
\eta^2}{4 \left(1-\eta\right)^3}-\frac{2-\eta}{1-\eta}G_3,
\label{n7b}
\eeq
{where we have made use of the definition of $G_0$, Eq.\ (\ref{6})}.

An extra condition is required to close the problem.  This follows
from the equal size limit given in Eq.\ (\ref{n2}), which after some
algebra yields
\beq
G_2=(2-\eta)g_{\text{p}}-\frac{2+\eta^2/4}{\left(1-\eta\right)^2},
\label{n6a}
\eeq
\beq
G_3=(1-\eta)\left(g_{\text{p}}^{\text{SPT}}-g_{\text{p}}\right),
\label{n6b}
\eeq
with
\beq
g_{\text{p}}^{\text{SPT}}=\frac{1-\eta/2+\eta^2/4}{(1-\eta)^3}
\label{SPT}
\eeq
 the contact value of the radial distribution function of the
one-component fluid in the SPT. It is also interesting to point out
that from Eqs.\ (\ref{n6a}) and (\ref{n6b}) it follows that
\beq
G_2+G_3=g_{\text{p}}-g_{\text{p}}^{\text{PY}},
\label{n8}
\eeq
where
\beq
g_{\text{p}}^{\text{PY}}=\frac{1+\eta/2}{(1-\eta)^2}
\label{PY}
\eeq
 is the contact value of the rdf for a one-component fluid in
the PY theory.

\begin{table*}[htb]
\begin{ruledtabular}
\begin{tabular}{cccccc}
Approximation &$g_\pure$& $G_1$ & $G_2$&
$Z-Z_w$ \\
\hline
PY&$\frac{1+\eta/2}{(1-\eta)^2}$&$\frac{3\eta}{2(1-\eta)^2}$&$0$&$\frac{3\eta^2\mt^3}{(1-\eta)^2\mth^2}$\\
SPT&$\frac{1-\eta/2+\eta^2/4}{(1-\eta)^3}$&$\frac{3\eta}{2(1-\eta)^2}$&$\frac{3\eta^2}{4(1-\eta)^3}$&0\\
BGHLL&$\frac{1-\eta/2}{(1-\eta)^3}$&$\frac{3\eta}{2(1-\eta)^2}$&$\frac{\eta^2}{2(1-\eta)^3}$&$\frac{\eta^2\mt^3}{(1-\eta)^2\mth^2}$\\
e1&Free&$g_\pure-\frac{1}{1-\eta}$&0&$\frac{2\mt^3}{\mth^2}\left[\frac{\mo\mth}{\mt^2}
({1-\eta})\left(g_\pure^{\text{PY}}-g_\pure\right)
+\eta\left(g_\pure-\frac{1}{1-\eta}\right)\right]$\\
e2&Free&$2\left(1-\eta\right)g_\pure-\frac{2-\eta/2}{1-\eta}$&$\frac{1-\eta/2}{1-\eta}-\left(1-2\eta\right)g_\pure$&
$\frac{4\mt^3}{\mth^2}\left(\frac{\mo\mth}{\mt^2}-
1\right){(1-\eta)^2}\left(g_\pure^{\text{SPT}}-g_\pure\right)$\\
VS&$\frac{1-\eta/2+\eta^3/4-\eta^4/2}{(1-\eta)^3}$&$\eta\frac{3-\eta+\eta^2/2}{2(1-\eta)^2}$&$\eta^2\frac{2-\eta-\eta^2/2}{3(1-\eta)^3}
\left(1+\frac{\mo\mth}{2\mt^2}\right)$&$\frac{\eta^2\mt^3}{6(1-\eta)^2\mth^2}\left[2+2\eta+7\eta^2-\frac{\mo\mth}{\mt^2}(2+5\eta-5\eta^2)\right]$
 \end{tabular}
\caption{Contact values of the one-component fluid $g_\pure$,
coefficients $G_1$ and $G_2$, and difference $Z-Z_w$ {corresponding
to} different theories.} \label{Tablec}
\end{ruledtabular}
\end{table*}

{In summary, from Eqs.\ (\ref{n1}), (\ref{7}), (\ref{5}),
 (\ref{n7a}), (\ref{n6a}), and (\ref{n6b}) we finally get the following expression
for the contact value of the particle-particle rdf:}
\beqa
g(\sigma,\sigma')&=&\frac{1}{1-\eta}+\frac{3 \eta}{
\left(1-\eta\right)^2}\frac{\mt}{\mth}\frac{\sigma
\sigma'}{\sigma+\sigma'}\nn
&&+4\left[(2-\eta)g_{\text{p}}-\frac{2+\eta^2/4}{\left(1-\eta\right)^2}\right]\left(\frac{\mt}{\mth}\frac{\sigma
\sigma'}{\sigma+\sigma'}\right)^2\nn
&&+8(1-\eta)\left(g_{\text{p}}^{\text{SPT}}-g_{\text{p}}\right)\left(\frac{\mt}{\mth}\frac{\sigma
\sigma'}{\sigma+\sigma'}\right)^3.
\label{e3}
\eeqa
{Similarly, the wall-particle expression is}
\beqa
g_w(\sigma)&=&\frac{1}{1-\eta}+\frac{3 \eta}{
\left(1-\eta\right)^2}\frac{\mt}{\mth}\sigma\nn
&&+4\left[(2-\eta)g_{\text{p}}-\frac{2+\eta^2/4}{\left(1-\eta\right)^2}\right]\left(\frac{\mt}{\mth}\sigma\right)^2\nn
&&+8(1-\eta)\left(g_{\text{p}}^{\text{SPT}}-g_{\text{p}}\right)\left(\frac{\mt}{\mth}\sigma\right)^3.
\label{e3w}
\eeqa
With the above results the compressibility factor may be finally
written in terms of $g_{\text{p}}$ as
\beqa
Z=Z_w&=&\frac{1}{\left(1-\eta\right)}+\frac{3 \eta}{
\left(1-\eta\right)^2}\frac{\mo \mt} {\mth}\nn &&+4 \eta
\frac{\mt^3}{\mth^2}\left[
g_{\text{p}}-\frac{1-\eta/4}{\left(1-\eta\right)^2}\right].
\label{ZZ}
\eeqa

 {Equations (\ref{e3})--(\ref{ZZ}) are the main
results of this paper.
 Note that the contact values of the system, and hence the EOS,  are wholly
determined once the EOS of the one-component fluid (and thus
$g_{\text{p}}$) is chosen. Therefore, our proposal  remains general
and flexible in the sense that, while fulfilling the consistency
conditions (\ref{n2}), (\ref{6}), and $Z=Z_w$, the choice of
$g_{\text{p}}$ can be done at will. Henceforth we will denote our
approximation by the label ``e3'' to emphasize that (i) it extends
any desired $g_{\text{p}}$ to the polydisperse case and (ii) $G(z)$
is a cubic function of $z$. When a particular one-component
approximation ``A'' is chosen, we will use the superscript ``eA3''
to refer to its extension. For instance, insertion of the
Carnahan--Starling EOS\cite{CS69}
\beq
g_{\text{p}}^{\text{CS}}=\frac{1-\eta/2}{(1-\eta)^3}
\label{CS}
\eeq
into Eqs.\ (\ref{e3}) and (\ref{e3w}) gives
$g^{\text{eCS3}}(\sigma,\sigma')$ and $g_w^{\text{eCS3}}(\sigma)$,
respectively. More specifically,}
\beqa
g^{\text{eCS3}}(\sigma,\sigma')&=&\frac{1}{1-\eta}+\frac{3 \eta}{
\left(1-\eta\right)^2}\frac{\mt}{\mth}\frac{\sigma
\sigma'}{\sigma+\sigma'}\nn
&&+\frac{\eta^2(1+\eta)}{(1-\eta)^3}\left(\frac{\mt}{\mth}\frac{\sigma
\sigma'}{\sigma+\sigma'}\right)^2\nn
&&+\frac{2\eta^2}{(1-\eta)^2}\left(\frac{\mt}{\mth}\frac{\sigma
\sigma'}{\sigma+\sigma'}\right)^3,
\label{eCS3}
\eeqa
\beqa
g_w^{\text{eCS3}}(\sigma)&=&\frac{1}{1-\eta}+\frac{3 \eta}{
\left(1-\eta\right)^2}\frac{\mt}{\mth}\sigma\nn
&&+\frac{\eta^2(1+\eta)}{(1-\eta)^3}\left(\frac{\mt}{\mth}\sigma\right)^2\nn
&&+\frac{2\eta^2}{(1-\eta)^2}\left(\frac{\mt}{\mth}\sigma\right)^3.
\label{eCS3w}
\eeqa

\begin{table}[htb]
\begin{ruledtabular}
\begin{tabular}{ccccc}
Mixture&Type&$\mt/\mo^2$&$\mth/\mo^3$ &$\eta$\\
\hline
1&Top-hat ($c=0.2$)&1.0133&1.04&0.2\\
2&Top-hat ($c=0.2$)&1.0133&1.04 &0.4\\
3&Top-hat ($c=0.7$)&1.1633&1.49&0.4\\
4&Schulz ($q=5$)&1.1667&1.5556&0.2\\
5&Schulz ($q=5$)&1.1667&1.5556&0.4
\end{tabular} \caption{Parameters of the size distributions for the examined mixtures.}
\label{TableM}
\end{ruledtabular}
\end{table}

\begin{table*}[htb]
\begin{ruledtabular}
\begin{tabular}{ccccccccccccc}
Mixture & MC & $Z^{\text{PY}}$ & $Z_w^{\text{PY}}$&
$Z^{\text{SPT}}$& $Z^{\text{BMCSL}}$ & $Z_w^{\text{BGHLL}}$ &
$Z^{\text{eCS1}}$& $Z_w^{\text{eCS1}}$& $Z^{\text{eCS2}}$& $Z_w^{\text{eCS2}}$ &$Z^{\text{VS}}$&$Z_w^{\text{VS}}$\\
\hline
 1&2.374&2.344& 2.163&2.389& 2.374& 2.314& 2.374& 2.240&2.374& 2.373&2.376&2.377\\
 2&6.746&6.197&4.915& 7.052&6.767& 6.340& 6.771& 5.636& 6.765 &6.762&6.792&6.743\\
 3&5.479&5.215&4.269& 5.845& 5.635& 5.320& 5.656& 4.848& 5.622& 5.603&5.663&5.642\\
 4&2.110&2.076&1.953& 2.107& 2.097& 2.056& 2.098& 2.012& 2.096&2.091&2.099& 2.102\\
 5&5.634&5.042&4.167& 5.625& 5.431&5.138& 5.458& 4.722&5.414&
 5.389&5.461&5.448
\end{tabular}
\caption{Comparison between the compressibility factor as obtained
by  MC simulations\protect\cite{Ignacio} and with the different
theories.}
\label{Z}
\end{ruledtabular}
\end{table*}
\subsection{Connection with former work}
As mentioned in the Introduction, the PY,\cite{L64} the
SPT,\cite{LHP65,R88} the BGHLL,\cite{B70,GH72,LL73} and our previous
approximations\cite{SYH99,Contact} for the contact values of the rdf
share the {universality} property indicated in Eq.\ (\ref{n1}).
{Furthermore, they have a polynomial dependence on $z$: linear in
the case of the PY approximation and the one we proposed in Ref.\
\onlinecite{SYH99} (here termed as ``e1''); quadratic in the case of
the SPT and BGHLL approximations, as well as in our quadratic
proposal of Ref.\ \onlinecite{Contact} (here termed as ``e2'').
Thus, these five approximations (of which only e1 and e2 allow for a
free choice of the one-component rdf $g_{\text{p}}$) may also be
expressed in the form of Eq.\ (\ref{5}), except that $G_3=0$. The
corresponding coefficients $G_1$ and $G_2$ appear in Table
\ref{Tablec}.} Further, we have also included a recent proposal by
Viduna and Smith (VS)\cite{VS02a} that may be cast into the form of
Eq.\ (\ref{5}) (again with $G_3=0$) but \textit{does not} comply
with the ansatz (\ref{n1}) since the coefficient $G_2$ depends on
the moments of the size distribution. Finally we have also included
a column with the difference between $Z$ and $Z_w$ for all {those}
theories.

\begin{figure}[htb]
\includegraphics[width=\columnwidth]{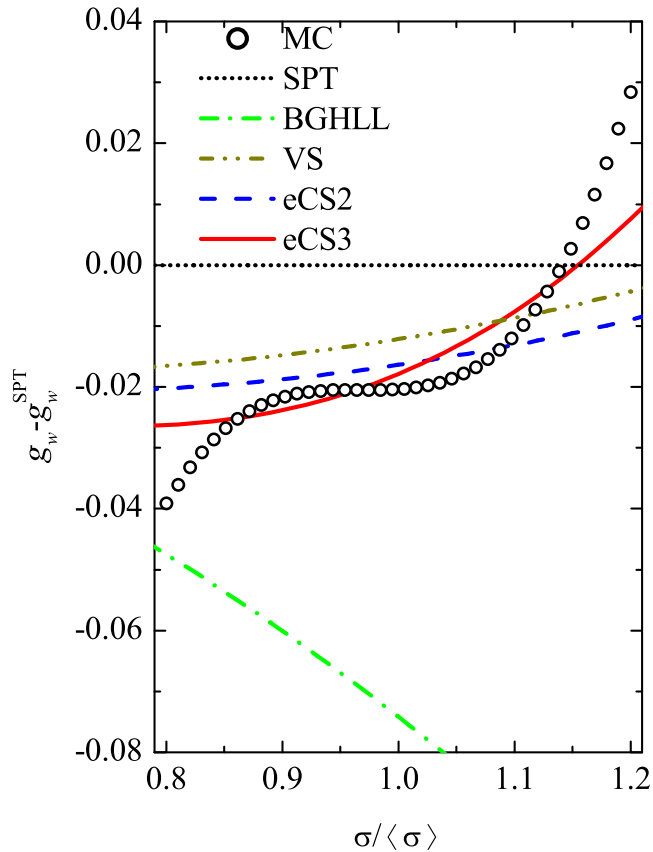}
\caption{(Color online). Plot of the difference of contact values
$g_w-g_w^{\text{SPT}}$
 as a function of $\sigma$
for a polydisperse mixture with a top-hat distribution
{(\protect\ref{th}) with $c=0.2$ at a packing fraction} $\eta=0.2$.
The symbols are MC simulations.\protect\cite{Ignacio} The lines are
SPT (\mbox{$\cdots$}),
 BGHLL (\mbox{-- $\cdot$ --
$\cdot$}), VS (\mbox{-- $\cdot\cdot$ -- $\cdot\cdot$}), eCS2
(\mbox{-- -- --}), and  eCS3 (\mbox{---}).
\label{fig1}}
\end{figure}
\begin{figure}[ht]
\includegraphics[width=.95\columnwidth]{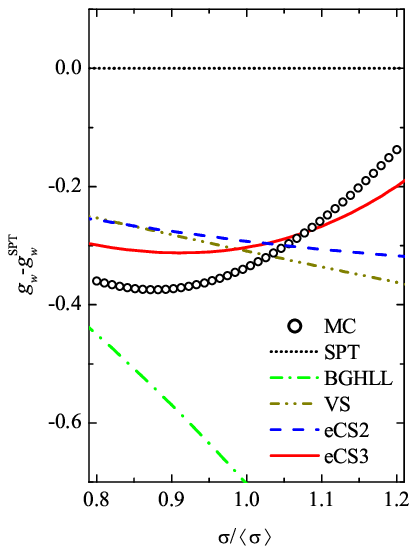}
\caption{(Color online). Plot of the difference of contact values
$g_w-g_w^{\text{SPT}}$
 as a function of $\sigma$
for a polydisperse mixture with a top-hat distribution
{(\protect\ref{th}) with $c=0.2$ at a packing fraction} $\eta=0.4$.
The symbols are MC simulations.\protect\cite{Ignacio} The lines are
SPT (\mbox{$\cdots$}),
 BGHLL (\mbox{-- $\cdot$ --
$\cdot$}), VS (\mbox{-- $\cdot\cdot$ -- $\cdot\cdot$}), eCS2
(\mbox{-- -- --}), and  eCS3 (\mbox{---}).\label{fig2}}
\end{figure}
\begin{figure}[ht]
\includegraphics[width=.95\columnwidth]{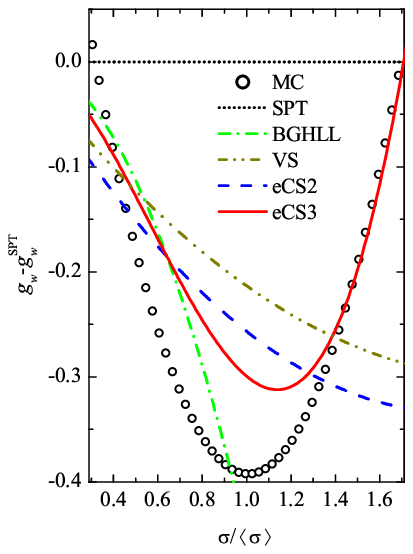}
\caption{(Color online). Plot of the difference of contact values
$g_w-g_w^{\text{SPT}}$
 as a function of $\sigma$
for a polydisperse mixture with a top-hat distribution
{(\protect\ref{th}) with $c=0.7$ at a packing fraction} $\eta=0.4$.
The symbols are MC simulations.\protect\cite{Ignacio} The lines are
SPT (\mbox{$\cdots$}),
 BGHLL (\mbox{-- $\cdot$ --
$\cdot$}), VS (\mbox{-- $\cdot\cdot$ -- $\cdot\cdot$}), eCS2
(\mbox{-- -- --}), and  eCS3 (\mbox{---}). \label{fig3}}
\end{figure}
\begin{figure}[ht]
\includegraphics[width=\columnwidth]{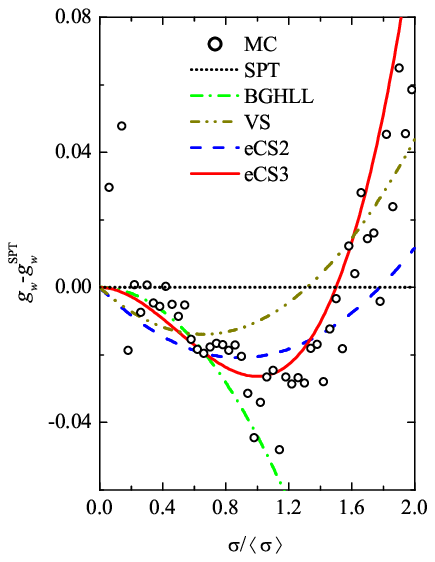}
 \caption{(Color online). Plot of the difference of contact values
$g_w-g_w^{\text{SPT}}$
 as a function of $\sigma$
for a polydisperse mixture with a Schulz distribution
{(\protect\ref{S}) with $q=5$ at a packing fraction} $\eta=0.2$. The
symbols are MC simulations.\protect\cite{Ignacio} The lines are SPT
(\mbox{$\cdots$}),
 BGHLL (\mbox{-- $\cdot$ --
$\cdot$}), VS (\mbox{-- $\cdot\cdot$ -- $\cdot\cdot$}), eCS2
(\mbox{-- -- --}), and  eCS3 (\mbox{---}).\label{fig4}}
\end{figure}
\begin{figure}[ht]
\includegraphics[width=\columnwidth]{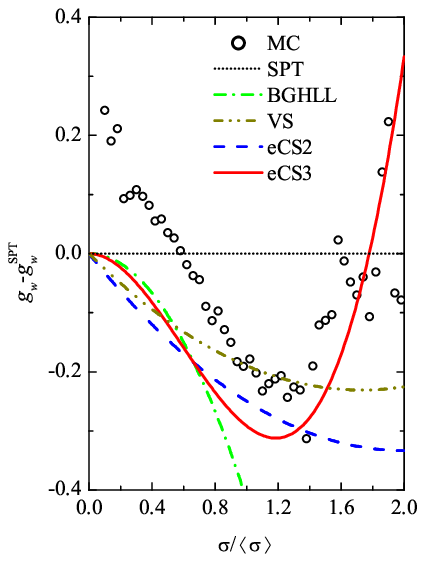}
 \caption{(Color online). Plot of the difference of contact values
$g_w-g_w^{\text{SPT}}$
 as a function of $\sigma$
for a polydisperse mixture with a Schulz distribution
{(\protect\ref{S}) with $q=5$ at a packing fraction} $\eta=0.4$. The
symbols are MC simulations.\protect\cite{Ignacio} The lines are SPT
(\mbox{$\cdots$}),
 BGHLL (\mbox{-- $\cdot$ --
$\cdot$}), VS (\mbox{-- $\cdot\cdot$ -- $\cdot\cdot$}), eCS2
(\mbox{-- -- --}), and  eCS3 (\mbox{---}). \label{fig5}}
\end{figure}
\begin{figure}[ht]
\includegraphics[width=.95\columnwidth]{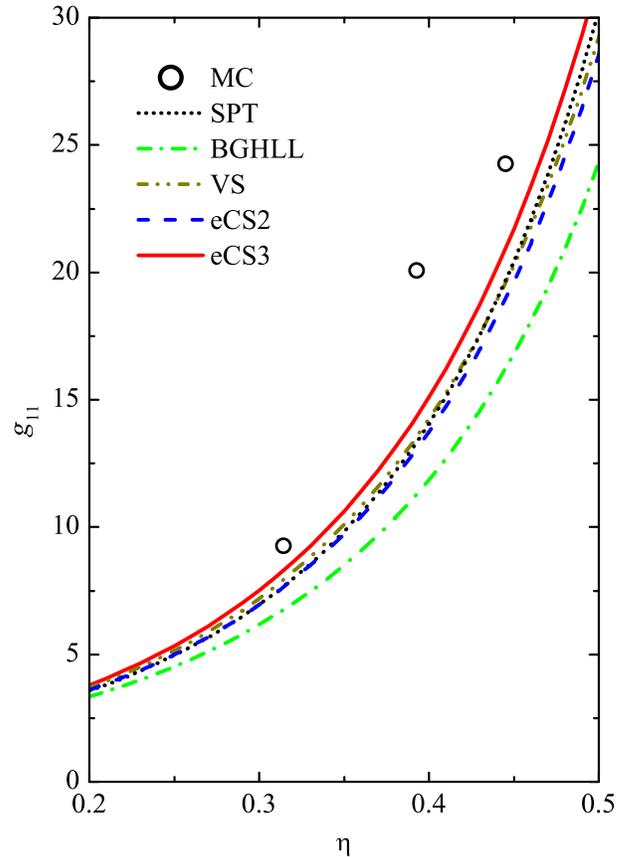}
 \caption{(Color online). Plot of the contact value $g_{11}(\sigma_{1})$ as a
 function of the packing fraction $\eta$ for the
binary hard-sphere mixture $x_1=0.005$, $\sigma_2/\sigma_1=0.2$.
 The symbols are MC simulations.\protect\cite{CCHW00}  The lines are
SPT (\mbox{$\cdots$}),
 BGHLL (\mbox{-- $\cdot$ --
$\cdot$}), VS (\mbox{-- $\cdot\cdot$ -- $\cdot\cdot$}), eCS2
(\mbox{-- -- --}), and  eCS3 (\mbox{---}). \label{fig6}}
\end{figure}
\begin{figure}[h]
\includegraphics[width=\columnwidth]{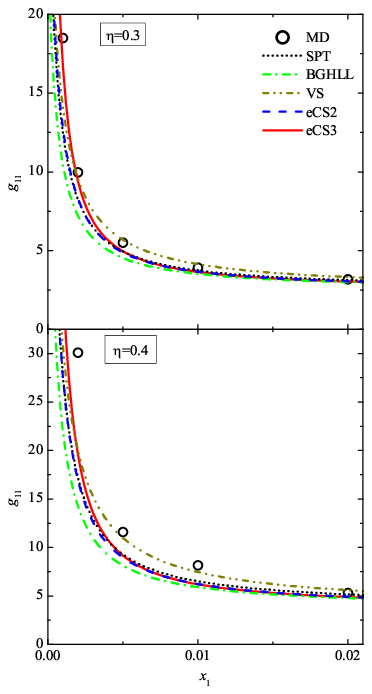}
\caption{(Color online). Plot of the contact value
$g_{11}(\sigma_{11})$ as a
 function of composition of the large spheres $x_1$ for two
binary hard-sphere mixtures with $\sigma_2/\sigma_1=0.1$ and packing
fractions $\eta=0.3$ (top panel) and $\eta=0.4$ (bottom panel),
respectively.
 The symbols are molecular dynamics (MD) simulations.\protect\cite{LW99,HTWC05}  The lines are
SPT (\mbox{$\cdots$}),
 BGHLL (\mbox{-- $\cdot$ --
$\cdot$}), VS (\mbox{-- $\cdot\cdot$ -- $\cdot\cdot$}), eCS2
(\mbox{-- -- --}), and  eCS3 (\mbox{---}).
 \label{fig7}}
 \end{figure}

From Table \ref{Tablec} we observe that, among the approximations
with $G_3=0$, only the SPT approximation yields a consistent EOS
through the virial and the hard-wall routes, for any density and any
degree of polydispersity. On the other hand, {this internal
consistency is at the expense of the rather poor quality of the SPT
contact value $g_\text{p}^{\text{SPT}}$ in the one-component case}.
The PY, BGHLL, e1 and VS approximations are not consistent {with
$Z=Z_w$}, even in the one-component limit ({in which case}
$\mn\to\mo^n$). {In the case of the e2 approximation,  the
inconsistency decreases with the degree of dispersity and disappears
in the one-component limit.}

{It must be noted that the e1 approximation embodies the PY
approximation as a particular case, \textit{i.e.}, the choice
$g_{\text{p}}=g_{\text{p}}^{\text{PY}}$ yields
$G^{\text{ePY1}}=G^{\text{PY}}$. Analogously, the SPT approximation
can be recovered from e2 and e3:
$g_{\text{p}}=g_{\text{p}}^{\text{SPT}}\Rightarrow
G^{\text{eSPT2}}=G^{\text{eSPT3}}=G^{\text{SPT}}$.}

{Another comment is in order at this stage. {}From Eq.\ (\ref{11})
we can observe that, for the class of approximations (\ref{5}), the
compressibility factor $Z$ does not depend on the individual values
of the coefficients $G_2$ and $G_3$, but only on their sum. As a
consequence, two different approximations of the form (\ref{5})
sharing the same density dependence of $G_1$ and $G_2+G_3$ also
share the same virial EOS. For instance, if one makes the choice
$g_{\text{p}}=g_{\text{p}}^{\text{PY}}$, then
$G^{\text{ePY3}}_2+G^{\text{ePY3}}_3=0$, see Eq.\ (\ref{n8}), and so
$Z^{\text{ePY3}}=Z^{\text{PY}}$, even though $G^{\text{ePY3}}(z)\neq
G^{\text{PY}}(z)$. Furthermore, if one makes the more sensible
choice $g_{\text{p}}=g_{\text{p}}^{\text{CS}}$, then
$G_{2}^{\text{eCS3}}+G_{3}^{\text{eCS3}}=G_{2}^{\text{BGHLL}}$, so
that $Z^{\text{eCS3}}=Z^{\text{BMCSL}}$, but again
$G^{\text{eCS3}}(z)\neq G^{\text{BGHLL}}(z)$. }

\section{Comparison with Monte Carlo simulations}
\label{sec3}

Buzzacchi  \textit{et al.}\cite{Ignacio} have recently computed by
Monte Carlo (MC) simulations {the wall-particle contact value}
$g_w(\sigma)$ and {the compressibility factor} $Z$ for polydisperse
hard spheres {of packing fractions $\eta=0.2$ and $\eta=0.4$} with
either a top-hat distribution of sizes given by
\beq
f(\sigma)=\begin{cases} 1/2c,& \mo(1-c)\leq\sigma\leq\mo(1+c)\\
0,&\text{otherwise}
\end{cases}
\label{th}
\eeq
or with a Schulz distribution of the form
 \beq
f(\sigma)=\frac{q+1}{q!\mo}\left(\frac{q+1}{\mo}\sigma\right)^q
\exp\left(-\frac{q+1}{\mo}\sigma\right).
\label{S}
\eeq
In Table \ref{TableM} we present the values of the parameters
corresponding to the examined mixtures.

Table \ref{Z} compares the MC data {of $Z$} for mixtures 1--5 with
{values obtained from different theoretical proposals by using both
the virial and the wall routes}. Here and it what follows, we have
made the choice $g_{\text{p}}=g_{\text{p}}^{\text{CS}}$ {in the
approximations e1, e2, and e3. As indicated above,
$Z^{\text{eCS3}}=Z_w^{\text{eCS3}}=Z^{\text{BMCSL}}$ and
$Z^{\text{SPT}}=Z_w^{\text{SPT}}$. Note that
$Z^{\text{BGHLL}}=Z^{\text{BMCSL}}$, but $Z_w^{\text{BGHLL}}\neq
Z^{\text{BMCSL}}$. It can be observed from Table \ref{Z} that (apart
from the SPT and the eCS3) the eCS2 and VS expressions for the
contact values provide the least internal inconsistency between $Z$
and $Z_w$.}  We can also observe that $Z^{\text{BMCSL}}$,
$Z^{\text{eCS1}}$, $Z^{\text{eCS2}}\simeq Z_w^{\text{eCS2}}$, and
$Z^{\text{VS}}\simeq Z_w^{\text{VS}}$ are the most accurate EOS for
{the top-hat} mixtures 1--3. On the other hand, the most accurate
EOS for {the Schulz} cases 4 and 5 is $Z^{\text{SPT}}$.

{Now we turn to the main topic of this paper, namely the contact
values of the rdf.} For the same polydisperse systems considered in
Table \ref{TableM}, in Figs.\ \ref{fig1}--\ref{fig5} we show  the
comparison between the simulation results\cite{Ignacio} for {the
wall-particle correlation function $g_w(\sigma)$}  and those of
different theories. Since all theories yield contact values which
are increasing functions of $\sigma$, in order to emphasize features
that would otherwise be difficult to ascertain, we have decided to
represent the difference $g_w-g_w^{\text{SPT}}$ rather than $g_w$.
The choice of $g_w^{\text{SPT}}$ as reference was motivated by the
fact that it is the only previous theory consistent with the two
ways of computing the pressure of the system. It is clear that the
best overall performance in the comparison with the simulation data
is given by $g_w^{\text{eCS3}}$, not only qualitatively but also
quantitatively. Interestingly enough, {$g_w^{\text{eCS2}}$ and
$g_w^{\text{VS}}$ also do a good job in the cases of mixtures 1
(Fig.\ \ref{fig1}) and 4 (Fig.\ \ref{fig4})}. In general, one can
see that the SPT overestimates the contact values, except for high
values of $\sigma$, and that the BGHLL prescription underestimates
them. Also, although not included in the figure to avoid
overcrowding, there is a very poor agreement of both
$g_w^{\text{eCS1}}$ and $g_w^{\text{PY}}$ (which are linear in $z$)
with the simulation data, both being underestimations.

While the main interest of our present formulation was geared
towards the polydisperse system near a hard wall, {Eq.\
(\ref{e3w})}, it should be clear that {our new proposal
 for the contact values of the rdf  also applies to the
bulk fluid, Eq.\ (\ref{e3})}. In particular,  if the diameter
distribution is discrete, {the replacements $\sigma\to\sigma_i$ and
$\sigma'\to\sigma_j$ in Eq.\ (\ref{e3}) yield} the contact values
$g_{ij}(\sigma_{ij})$, where $i$ and $j$ denote the $i$-th and
$j$-th species, respectively, and $\sigma_{ij}\equiv
\left(\sigma_{i}+\sigma_{j}\right)/2$, with $\sigma_{k}$  denoting
the diameter of a sphere of species $k$. Just to illustrate the kind
of results our proposal produces, we consider two examples of binary
hard-sphere mixtures. In the first case, we show in Fig.\ \ref{fig6}
a plot of the different theoretical predictions of
$g_{11}(\sigma_{1})$ as a
 function of the packing fraction $\eta$, for a
mixture having a mole fraction of the large spheres $x_1=0.005$, and
a size ratio $\sigma_2/\sigma_1=0.2$, together with the simulation
results of Cao \textit{et al.}\cite{CCHW00} On the other hand, in
Fig.\ \ref{fig7} and for a binary mixture with
$\sigma_2/\sigma_1=0.1$ and two values of the packing fraction
($\eta=0.3$ and $\eta=0.4$), we display the behavior of
$g_{11}(\sigma_{1})$ as a function of the mole fraction of species
$1$ derived from the various theories and the  results of simulation
by Lue and Woodcock\cite{LW99}  and Henderson \textit{et
al.}\cite{HTWC05} It is again clear from Fig. \ref{fig6} that
$g_{11}^{\text{eCS3}}(\sigma_{1})$ gives the best performance for
this mixture. As far as the behavior with respect to the mole
fraction of the large spheres is concerned, as already noted by
Henderson \textit{et al.},\cite{HTWC05} the approximation of Viduna
and Smith\cite{VS02a} does a very good job, especially for
$\eta=0.4$, but it goes wrong for small values of $x_1$. Figure
\ref{fig7} also indicates that our proposal is certainly the best
for $\eta=0.3$ and that it also accounts correctly for the sharp
rise observed at the smallest values of the mole fraction of species
$1$ for both packing fractions.

\section{Concluding remarks}
\label{sec4}

In this paper we have provided a new proposal for the contact values
of the particle-particle correlation function, $g(\sigma,\sigma')$,
and  of the wall-particle correlation function, $g_w(\sigma)$,  of a
hard-sphere fluid mixture with an arbitrary size distribution. The
proposal relies on a kind of universality assumption by which, once
the packing fraction is fixed, for all pairs of like and unlike
spheres the dependence of the contact values  on the diameters and
on the composition is only through a single dimensionless parameter
and holds for an arbitrary number of components. It also makes use
of the point particle and the equal size limits and of the
{internal} consistency between the usual virial route and the
hard-wall limit route to derive the pressure of the system. As a
consequence, the contact value $g_{\text{p}}$ of the rdf of a
one-component fluid is required as the only input, thus making the
formulation to be both simple and rather flexible.

The merits of this proposal have been assessed by comparing the
contact values themselves and the corresponding compressibility
factors with other theoretical developments and with recent MC
simulation results both for polydisperse hard-sphere fluids at a
hard wall and for binary hard-sphere mixtures (discrete size
distribution). It is fair to say that the new proposal with the CS
expression for $g_{\text{p}}$ gives the best overall performance.
Also it is clear that {(i) two different approximations
$g(\sigma,\sigma')$ for the contact values can yield the same
compressibility factor $Z$ and (ii) a fortunate cancelation of
errors can make a poor approximation for $g(\sigma,\sigma')$ to lead
to a reliable $Z$. Examples of the first effect are provided by the
approximations PY and ePY3, which differ at the level of the contact
values but share the same EOS, and, similarly, by the approximations
BGHLL and eCS3. An example of the second effect is represented by
the eCS1 approximation, which yields a very accurate
EOS,\cite{SYH99} even though the associated contact values are only
qualitatively correct.}

In previous work of ours we have attempted to provide expressions
for the contact values of the rdf and the compressibility factors
that are valid for any dimensionality $d$.\cite{SYH99,Contact} The
present proposal does not fulfill such a condition. It will also
work for $d=1$ and $d=2$,  but it is not {directly} generalizable to
arbitrary $d$.\cite{LS04}

Apart from the EOS, contact values of the rdf may also be useful in
other contexts. For instance, they are required as input in the
rational function approximation approach to the structural
properties of hard-sphere mixtures.\cite{YSH98} We plan to use them
in connection with this problem in the near future.

\acknowledgments

We want to thank M. Buzzacchi, I. Pagonabarraga, and N. B. Wilding
for kindly providing us with tables of their numerical MC results.
The research of A.S. and S.B.Y. has been supported by the Ministerio
de Educaci\'on y Ciencia (Spain) through grant No.\ FIS2004-01399
(partially financed by FEDER funds).


\begin{thebibliography}{99}
\bibitem{LS04}
{For $d=1$, the approximations e1, e2, and e3 reduce to the exact
result $g(\sigma,\sigma')=1/(1-\eta)$. For $d=2$, the approximations
e3 and e2 become identical because the latter already fulfills the
condition $Z=Z_w$, as shown by S. Luding and A. Santos,  J. Chem.
Phys. \textbf{121}, 8458 (2004). It is tempting to speculate  that a
polynomial form for $G(z)$ of degree $d$ could be found to be
consistent with the condition $Z=Z_w$ for $d>3$. However, a more
detailed analysis shows that this is not the case since the number
of conditions exceeds the number of unknowns, unless the
universality assumption is partially relaxed.}
\end{thebibliography}

\end{document}